\documentclass[aps,twocolumn,prb,showpacs]{revtex4}

\usepackage[dvips]{color}
\usepackage{graphicx}
\usepackage{epsfig}
\usepackage{dcolumn}
\usepackage{amssymb}
\usepackage[english]{babel}
\usepackage{here}
\usepackage{mathrsfs}
\usepackage{bm}
\usepackage{amsmath}
\usepackage{lscape}

\newcommand{\KK}{\mathbf{K}}
\newcommand{\kk}{\mathbf{k}}            
\newcommand{\GG}{\mathbf{G}}            
\newcommand{\mat}[1]{[\![#1]\!]}        

\begin{document}

\title{Quantum theory of exciton-photon coupling in photonic crystal
slabs \\ with embedded quantum wells }

\author{Dario Gerace$^{*}$}
\affiliation{Dipartimento di Fisica ``Alessandro Volta,''
Universit\`a di Pavia, Via Bassi 6, I-27100 Pavia, Italy \\
and Institute of Quantum Electronics, ETH Zurich, 8093 Zurich,
Switzerland}

\author{Lucio Claudio Andreani}
\affiliation{Dipartimento di Fisica ``Alessandro Volta,''
Universit\`a di Pavia, Via Bassi 6, I-27100 Pavia, Italy}

\date{\today}

\begin{abstract}
A theoretical description of radiation-matter coupling for
semiconductor-based photonic crystal slabs is presented, in which
quantum wells are embedded within the waveguide core layer. A full
quantum theory is developed, by quantizing both the
electromagnetic field with a spatial modulation of the refractive
index and the exciton center of mass field in a periodic piecewise
constant potential. The second-quantized hamiltonian of the
interacting system is diagonalized with a generalized Hopfield
method, thus yielding the complex dispersion of mixed
exciton-photon modes including losses. The occurrence of both weak
and strong coupling regimes is studied, and it is concluded that
the new eigenstates of the system are described by quasi-particles
called \textit{photonic crystal polaritons}, which can occur in
two situations: (\textit{i}) below the light line, when a
resonance between exciton and non-radiative photon levels occurs
(\textit{guided polaritons}), (\textit{ii}) above the light line,
provided the exciton-photon coupling is larger than the intrinsic
radiative damping of the resonant photonic mode (\textit{radiative
polaritons}). For a square lattice of air holes, it is found that
the energy minimum of the lower polariton branch can occur around
normal incidence. The latter result has potential implications for
the realization of polariton parametric interactions in photonic
crystal slabs.
\end{abstract}



\maketitle

\section{Introduction}\label{intro}

The capability of engineering electron and photon states through
spatial confinement leads to a control of radiation-matter
interaction and to a number of interesting results in the field of
solid-state cavity quantum electrodynamics
(QED).\cite{varenna_proc}  For example, the use of quantum-well
(QW) excitons embedded in high-finesse semiconductor microcavities
(MC) of the Fabry-P\'erot type has allowed to observe a
modification of spontaneous emission (weak coupling
regime)\cite{bjork91} as well as the occurrence of a vacuum Rabi
splitting (strong coupling regime).\cite{weisbuch92,houdre94a} The
latter effect arises when the radiation-matter coupling energy
overcomes the damping rates of QW excitons and MC photons. Under
such conditions, the elementary excitations of the system should
not be described as barely excitonic or photonic, but rather as
mixed radiation-matter states, called \textit{MC
polaritons}.\cite{skolnik98,khitrova99,savona_cargese,kavokin_book,andreani_sif}
The physics of exciton-polaritons in bulk materials had been
introduced in the fifties \cite{hopfield58,agranovich} and is now
a textbook topic.\cite{klingshirn_book,cardona_book} The
confinement of both excitons and photons with MC-embedded QWs
leads to polariton states that are more easily observed, even at
room temperature, and in fact MC polaritons have been a major
research topic in the nineties. In 2000, the demonstration of
coherent parametric processes due to polariton-polariton
scattering\cite{savvidis00,saba01nat} opened a new area of
research in the physics of MC polaritons and nonlinear optics.
More recently, the experimental demonstration of Bose-Einstein
condensation of MC polaritons with II-VI
materials\cite{kasprzak06nat} brings a new exciting development
and demonstrates the potentialities of radiation-matter
interaction in the presence of electron and photon confinement.

Until now, little has been done to study radiation-matter
interaction when high-quality QWs are combined with new routes for
the tailoring of electromagnetic field in photonic crystals (PhC).
These structures, which are characterized by a periodic modulation
of the dielectric constant along one, two or three dimensions (1D,
2D, 3D),\cite{yablo87} allow for an unprecedented control over the
propagation and confinement of light at optical wavelengths. In
particular, two-dimensional PhC embedded in a planar dielectric
waveguide, commonly known as \textit{photonic crystal slabs}, are
receiving much attention because they allow for a 3D control of
light and are easily realized at sub-micron length
scales.\cite{sakoda_book,johnson_book,busch_book,lourtioz_book} In these
systems, light propagation or confinement is controlled by the PhC
structure in the 2D plane and by the dielectric discontinuity
provided by the slab waveguide in the vertical direction. When QWs
are embedded in a PhC slab, additional flexibility and new effects
related to exciton-photon coupling are to be expected.

In this work we analyze radiation-matter interaction in PhC slabs,
by considering the effects of coupling between QW excitons and
photonic modes. A main goal is to study the conditions leading to
the strong coupling regime and to point out the possibilities
offered by PhC for \textit{strong coupling engineering}. In the
past, both experimental\cite{fujita98} and
theoretical\cite{yablonskii01} investigations of strong coupling
regime in organic-based 1D PhC slabs have been reported, the
theory being based on a semiclassical treatment. The same groups
also reported a joint work on 2D PhC slabs with organic active
medium, in which the theoretical treatment was again based on a
semiclassical solution of Maxwell's equations within a scattering
matrix formalism.\cite{shimada02ieee} Exciton-polaritons in 1D
PhCs (ideal Bragg mirrors) in which one of the constituent media
has a strong excitonic character were theoretically studied by
using a fully classical approach.\cite{nojima98} Other theoretical
works focused on polaritonic gaps in so-called phonon-polaritonic
PhCs,\cite{huang03} or on the properties of 3D arrays of
quasi-zero-dimensional excitons (e.g., quantum dots) coupled to
Bloch modes of the same periodic array of dielectric
objects.\cite{ivchenko00} More recently, a purely 2D square
lattice array of rods in a dielectric host medium has been
theoretically studied, in which one of the two constituent media
has a frequency-dependent permittivity with a pole at a resonant
frequency.\cite{poddubny07}
Combining the effects of exciton confinement in QW structures with
the 2D periodicity of the surrounding dielectric environment has
been only studied in relation to the influence of Coulomb
interaction on the longitudinal part of the electromagnetic field
for deeply patterned PhCs with an unpatterned QW underneath the PhC
region.\cite{eichman03} No experimental or theoretical work has been
reported so far, to the best of our knowledge, on excitons or
exciton-polaritons in semiconductor-based PhC structures that are
the subject of this work.

In the present paper, the theoretical problem of exciton-photon
coupling in PhC slabs is tackled by using a fully
quantum-mechanical formalism for both photons and exciton states,
which is described in detail. With respect to previous works on
the quantum theory of exciton-polaritons in semiconductor
nano-structures,\cite{savona94,jorda94,citrin94,pau95prb,savasta96,panzarini99}
the non-trivial spatial dependence of the dielectric constant is
taken into account by quantization of the electromagnetic field in
a non-homogeneous
medium.\cite{carniglia71,glauber91pra,ho_cargese,sipe06pra} The
total hamiltonian for the coupled exciton-photon states is derived
and diagonalized numerically to obtain the eigenenergies of the
mixed modes. It is shown that polaritonic effects are typically
stronger in PhC slabs than in MCs, due to the better field
confinement provided by total internal reflection in the slab. As
a consequence, larger vacuum Rabi spitting is found at
exciton-photon resonance in strong coupling regime, which occurs
when the exciton-photon coupling energy is larger than the
intrinsic radiative linewidth of the photonic mode and the QW
exciton. In this scenario, new quasi-particles form in the PhC
slab, which we call \textit{photonic crystal polaritons}.
Preliminary results of the formalism presented here have been
published as conference
proceedings,\cite{gerace04pss,andreani04pn,andreani05pss} and
limited to 1D or triangular PhC lattices. In addition to the full
theoretical formulation, we present here new results for a PhC
slab made of a square lattice of air holes in a high-index
suspended membrane. For a specific PhC slab design, a polariton
angular dispersion is found which has a minimum close to normal
incidence with respect to the slab surface. Such result paves the
way for new exciting developments in the investigation of hybrid
semiconductor structures exhibiting both photonic and electronic
band gap characteristics, as it will be discussed.

The paper is organized as follows. In Sec.~\ref{system} we
schematically describe the system under investigation. In
Sec.~\ref{theory} we give a detailed account of the
second-quantization procedure leading to the complete
exciton-photon hamiltonian in the linear (low excitation density)
regime and of the diagonalization technique. In Sec.~\ref{results}
we provide systematic results in both weak and strong coupling
regimes, and describe the formation of guided and radiative
polaritons. The relevance of the results in the context of
parametric processes involving polariton-polariton scattering is
also discussed. Finally, in Sec.~\ref{concl} we summarize the
conclusions of the work.

\begin{figure}[t]
\begin{center}
\includegraphics[width=0.45\textwidth]{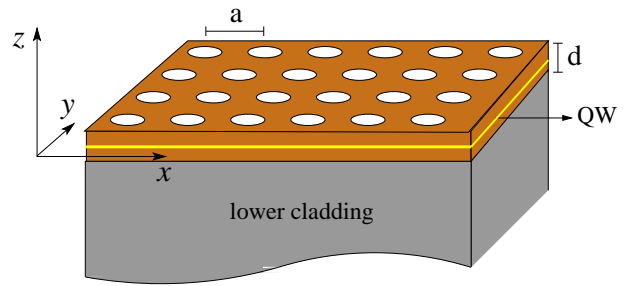}
\caption{(color online) Schematic view of a two-dimensional
photonic crystal slab of thickness $d$ and lattice constant $a$,
with a single quantum well grown within the core
layer.\label{fig1} }
\end{center}
\end{figure}

\section{The system}\label{system}

The schematic of the model system under investigation and the
choice of the coordinate axes are given in Fig.~\ref{fig1}. We
consider a high-index planar dielectric waveguide with
semi-infinite cladding layers (typically the upper cladding is
air). We assume that a QW of thickness $L_{\mathrm{QW}}$ is grown
in the core of the waveguide at a certain vertical position
$z_{\mathrm{QW}}$, measured from the core/lower cladding
interface. The dielectric material constituting the core of the
planar waveguide acts also as a barrier material for the carriers
(electrons and holes) confined in the thin QW layer. Typical QW
thickness is of the order of $L_{\mathrm{QW}}\simeq 10$ nm, while
the dielectric core layer is usually between $d=100$ and 200 nm
thick. The QW exciton is characterized by its transition energy,
$E_{\mathrm{ex}}$, and by the oscillator strength per unit area,
$f/S$.\cite{andreani_erice} The effect of multiple quantum wells
(MQWs) grown in the core layer can be taken into account by a
total oscillator strength given by the sum of single QW oscillator
strengths. This approach is valid as long as $L_{\mathrm{QW}}\ll
d$, and the electric field of the relevant photonic mode can be
considered as uniform along the MQW thickness. We disregard the
difference in dielectric constants between QW and barrier media,
and assume the core layer to be described by a relative dielectric
permeability $\epsilon_{\mathrm{diel}}$. The core layer is
patterned with a given photonic lattice, e.g. with lattice
constant $a$, down to the interface with the lower cladding
medium. Thus, exciton center-of-mass eigenfunctions are not free
as in usual QWs, but are subject to a further confining potential
provided by the etched air regions. As a general remark, it is
clear that exciton and photon wavefunctions are confined in the
vertical direction and are subject to effective potentials having
the same spatial periodicity in the 2D plane. Roughly speaking, we
could say that confined excitons and photons display the same
dimensionality in this problem, thereby satisfying the condition
for possible occurrence of quasi-stationary, strongly coupled
polariton states.\cite{andreani_sif,note_dimensionality}

\section{Quantum theory of exciton-photon coupling}\label{theory}

In this Section, we detail the full quantum theory of
radiation-matter interaction for QWs embedded in a PhC slab. We
start by the second quantization of the electromagnetic field in a
non-uniform dielectric environment, to which end we follow the
treatment given in Ref.~\onlinecite{glauber91pra}. Then, we
quantize the exciton field by solving the Schr\"{o}dinger equation
for the center-of-mass QW exciton wavefunction in a periodic,
piecewise constant potential. The solutions for the
non-interacting fields are used to derive the coupling
hamiltonian. The total hamiltonian is then diagonalized, leading
to PhC polariton eigenmodes.

\subsection{Quantized photon field in non-uniform dielectric medium:
normal mode expansion}\label{subsec:emfield}

The canonical quantization of the electromagnetic field in a
non-uniform dielectric medium with relative permeability
$\epsilon(\mathbf{r})$ (we assume a unit magnetic permeability) is
carried out through quantization of the vector potential
$\mathbf{A}$, defined from the usual relations to the fields
\begin{eqnarray}
\mathbf{E}(\mathbf{r},t)&=&-\frac{1}{c}\frac{\partial\mathbf{A}(\mathbf{r},t)}{\partial
t}=-\frac{1}{c}\dot{\mathbf{A}}(\mathbf{r},t)\, ,\\
\mathbf{B}(\mathbf{r},t)&=&\nabla\times\mathbf{A}(\mathbf{r},t) \, ,
\end{eqnarray}
and which must satisfy the equation of motion
\begin{equation}
\mathbf{\nabla}\times\mathbf{\nabla}\times\mathbf{A}
=\frac{\omega^2}{c^2} \epsilon(\mathbf{r})\mathbf{A}
\,\label{eq:maxwell2}.
\end{equation}
We are considering only the retarded electromagnetic field (no
scalar potential, $\Phi=0$) and the vector potential can be chosen
to satisfy the generalized Coulomb gauge\cite{glauber91pra}
\begin{equation}
\nabla\cdot(\epsilon(\mathbf{r})\mathbf{A}(\mathbf{r},t))=0
\,\label{eq:gauge}.
\end{equation}
The second-quantized hamiltonian of the free photon field is
straightforwardly obtained from the classical expression of the
total electromagnetic energy
\begin{equation}
\mathcal{H}_{\mathrm{e.m.}}= \frac{1}{8\pi}\int_V
\left[\epsilon(\mathbf{r})\mathbf{E}(\mathbf{r})^2 +
\mathbf{B}(\mathbf{r})^2\right]\,
\mathrm{d}\mathbf{r}
\,\label{elmagnen},
\end{equation}
which can be derived from the Proca Lagrangian
density\cite{cohen_book} and can be expressed in terms of
$\mathbf{A}(\mathbf{r},t)$ and its conjugate momentum
$\Pi(\mathbf{r},t)=\epsilon(\mathbf{r})\dot{\mathbf{A}}(\mathbf{r},t)/c^2$.
The field operator is expanded in normal modes as
\begin{widetext}
\begin{equation}
\hat{\mathbf{A}}(\mathbf{r},t)=\sum_{\kk,n}
\left(2\pi\hbar\omega_{\kk n}\right)^{1/2} \left[\hat{a}_{\kk
n}\mathbf{A}_{\kk n}(\mathbf{r}) e^{-i\omega_{\kk
n}t}+\hat{a}_{\kk n}^{\dagger} \mathbf{A}_{\kk
n}^{\ast}(\mathbf{r}) e^{i\omega_{\kk
n}t}\right]
\,\label{sqAfield},
\end{equation}
\end{widetext}
in which $\hat{a}_{\kk n}^{\dagger}$ ($\hat{a}_{\kk n}$) are
creation (destruction) operators of field quanta with
eigenenergies $\omega_{\kk n}$, and the index $n$ is a generic
band number labelling the corresponding photonic eigenmode at
in-plane Bloch vector $\kk$. In order to satisfy Bose commutation
relations for the operators,
\begin{equation}
[\hat{a}_{\kk n},\hat{a}_{\kk' n'}^{\dagger}]=
\delta_{\kk,\kk'}\delta_{n,n'}\,,\, [\hat{a}_{\kk n},\hat{a}_{\kk'
n'}]= [\hat{a}_{\kk n}^{\dagger},\hat{a}_{\kk' n'}^{\dagger}]=0
\,\label{commrel},
\end{equation}
the correct orthonormality condition for the classical functions
$\mathbf{A}_{\kk n}(\mathbf{r})$ satisfying Eq.
(\ref{eq:maxwell2}) is\cite{glauber91pra,carniglia71,sipe06pra}
\begin{equation}
\int_{V}\epsilon(\mathbf{r})\mathbf{A}_{\kk
n}^{\ast}(\mathbf{r})\cdot \mathbf{A}_{\kk' n'}(\mathbf{r})
\mathrm{d}\mathbf{r}= \frac{c^2}{\omega_{\kk n}^2}
\delta_{\kk,\kk'}\delta_{n,n'}
\,\label{normAfield},
\end{equation}
where $V$ is the same quantization volume as in (\ref{elmagnen}).
This implies the following conditions for the electric and
magnetic fields:
\begin{eqnarray}
\int_{V}\epsilon(\mathbf{r})\mathbf{E}_{\kk
n}^{\ast}(\mathbf{r})\cdot \mathbf{E}_{\kk' n'}(\mathbf{r})
\mathrm{d}\mathbf{r}= \delta_{\kk,\kk'}\delta_{n,n'}\,\label{normEfield}, \\
\int_{V}\mathbf{B}_{\kk n}^{\ast}(\mathbf{r})\cdot
\mathbf{B}_{\kk' n'}(\mathbf{r}) \mathrm{d}\mathbf{r}=
\delta_{\kk,\kk'}\delta_{n,n'}
\,\label{normBfield}.
\end{eqnarray}

The usual Maxwell equations for $\mathbf{E}$ and $\mathbf{B}$ can
be derived from the equations of motion for the conjugate
variables $\mathbf{A}(\mathbf{r})$ and $\mathbf{\Pi}(\mathbf{r})$.
From Eq.~(\ref{elmagnen}) and using the field expansion
(\ref{sqAfield}), we finally get the second-quantized hamiltonian
of the free photon field in the usual form as a sum of harmonic
degrees of freedom,
\begin{equation}
\hat{H}_{\mathrm{ph}}=\sum_{\kk,n}\hbar\omega_{\kk n}
\left(\hat{a}_{\kk n}^{\dagger} \hat{a}_{\kk
n}+\frac{1}{2}\right)
\,\label{photham}.
\end{equation}

In order to implement the formalism, we shall need specific forms
for the classical fields functions. The exact solution of Maxwell
equations in a PhC slab is a complicated task, especially
concerning quasi-guided modes that lie above the light
line.\cite{sakoda_book,johnson_book} A recently developed
guided-mode expansion (GME) approach\cite{andreani06prb} allows us
to find a convenient solution to this problem after expansion of
the classical fields on the basis of guided modes of an effective
planar waveguide. We start from the second-order equation for the
magnetic field $\mathbf{B}\equiv\mathbf{H}$ in a source-free
dielectric medium and for harmonic time dependence,
\begin{equation}
\mathbf{\nabla}\times\left[ \frac{1}{\epsilon(\mathbf{r})}
\mathbf{\nabla}\times\mathbf{H} \right] =\frac{\omega^2}{c^2}
\mathbf{H}
\, \label{eq:maxwell}.
\end{equation}
Due to the in-plane translational invariance implying
Bloch-Floquet theorem, the magnetic field can be expanded on a
basis in which planar and vertical coordinates are factorized
\begin{equation}
\mathbf{H}_{\kk}(\mathbf{r}) =  \sum_{\GG}\sum_{\alpha}
c_{\kk,\alpha}(\GG) \mathbf{h}_{\kk+\GG,\alpha}(z)
\mathrm{e}^{i(\mathbf{k}+\mathbf{G})
\cdot{\mbox{\boldmath\scriptsize$\rho$}}}
\,\label{eq:expansion},
\end{equation}
where $\mathbf{r}=({\mbox{\boldmath$\rho$}},z)$, $\kk$ is the
in-plane Bloch vector in the first Brillouin zone (BZ), $\GG$ are
reciprocal lattice vectors, and the functions
$\mathbf{h}_{\kk+\GG,\alpha}(z)$ ($\alpha=1,2,...$) are the
(discrete) guided modes of the effective planar waveguide with an
average dielectric constant in each layer, calculated from the air
fraction of the given photonic lattice. Thus,
Eq.~(\ref{eq:maxwell}) is reduced to a linear eigenvalue problem
\begin{equation}
\sum_{\GG'}\sum_{\alpha'}
\mathcal{M}_{\kk+\GG,\kk+\GG'}^{\alpha,\alpha'}
c_{\alpha'}(\kk+\GG')=\frac{\omega^2}{c^2}c_{\alpha}(\kk+\GG)
\,\label{eq:linear},
\end{equation}
which is solved by diagonalizing the explicit expression for the
matrix $\mathcal{M}$.\cite{andreani06prb} The properties of the
specific photonic lattice enter in the matrix $\mathcal{M}$ as a
Fourier transform of the inverse dielectric constant in each
layer, $\eta(\GG,\GG')=\epsilon^{-1}(\GG,\GG')$, the matrix
inversion being performed numerically. After diagonalization, the
resulting photonic modes can be classified according to their band
index, $n$, and their in-plane Bloch vector $\kk$. It should be
noted that the classical fields $\mathbf{H}_{\kk n}(\mathbf{r})$
and $\mathbf{E}_{\kk
n}(\mathbf{r})=ic/(\omega\epsilon(\mathbf{r}))\nabla\times\mathbf{H}_{\kk
n}(\mathbf{r})$ calculated by the GME approach automatically
satisfy the orthonormality conditions
(\ref{normEfield}-\ref{normBfield}), thus they constitute a very
convenient set for the second-quantized formulation. Finally, we
point out that losses of a quasi-guided photonic eigenmode can be
introduced in the present treatment as an imaginary part
$\gamma_{\kk n}$ of mode eigenenergies. Such imaginary part arises
for modes lying above the light line from out-of-plane diffraction
losses, which are calculated in perturbation theory as detailed in
Ref.~\onlinecite{andreani06prb}.

\subsection{Quantized exciton field in a periodic piecewise
constant potential}\label{subsec:exfield}

Exciton confinement in QWs has been widely investigated in the
past (for a review see, e.g., Ref.~\onlinecite{andreani_erice}).
Also, second-quantization of the QW exciton field has been
extensively treated in the literature. Two recent approaches
leading to an effective hamiltonian of interacting bosons starting
from the full crystal hamiltonian are given in
Refs.~\onlinecite{tassone99,rochat00}. When the exciton density
$n_{\mathrm{ex}}$ is low, i.e., much smaller than a saturation
density $n_{\mathrm{sat}}\simeq 1/(2\pi a^2_{2D})$ (where $a_{2D}$
is the 2D exciton Bohr radius),\cite{ciuti00,ciuti03sst} QW
excitons behave as a gas of non-interacting bosons to a very good
approximation.

The envelope function
$F(\mathbf{r}_{\mathrm{e}},\mathbf{r}_{\mathrm{h}})$ of QW
excitons can be factorized in a wavefunction depending on a
center-of-mass coordinate $\mathbf{R}_{\parallel}$ and an
electron-hole wavefunction depending on the in-plane relative
coordinate ${\mbox{\boldmath$\rho$}}$ as
\begin{equation}
F(\mathbf{r}_{\mathrm{e}},\mathbf{r}_{\mathrm{h}})=
F_{\KK}(\mathbf{R}_{\parallel})
f({\mbox{\boldmath$\rho$}},z_{\mathrm{e}},z_{\mathrm{h}})
\,\label{envelopeqw},
\end{equation}
where $F_{\KK}(\mathbf{R}_{\parallel})=e^{i\mathbf{K}\cdot
\mathbf{R}_{\parallel}}$ for free QW excitons. Considering the
system of Fig.~\ref{fig1}, the QW layer is patterned with a
photonic lattice of air holes and the center-of-mass wavefunction
$F_{\KK}(\mathbf{R}_{\parallel})$ for exciton motion in the 2D
plane is not a simple plane wave anymore.\cite{note_dead_layer}
This wavefunction obeys a Schr\"{o}dinger equation
\begin{equation}
\left[-\frac{\hbar^2 \nabla^2} {2M_{\mathrm{ex}}}
+V(\mathbf{R}_{\parallel})\right]
\,F_{\KK}(\mathbf{R}_{\parallel})=
E_{\KK}\,F_{\KK}(\mathbf{R}_{\parallel})
\,\label{eq:schrodinger},
\end{equation}
where
$M_{\mathrm{ex}}=m_{\mathrm{e}}^{\ast}+m_{\mathrm{h}}^{\ast}$ is
the total exciton mass and the potential
$V(\mathbf{R}_{\parallel})=0$ in the non-patterned regions, while
$V(\mathbf{R}_{\parallel})$ takes a large value $V_{\infty}$ in
the air holes. Equation~(\ref{eq:schrodinger}) can be solved by
plane-wave expansion,
\begin{equation}
F_{\KK}(\mathbf{R}_{\parallel}) = \sum_{\GG} F(\KK+\GG)
\mathrm{e}^{i(\KK+\mathbf{G})\cdot{\mathbf{R}_{\parallel}}}
\,\label{eq:excexp},
\end{equation}
where the exciton wavevector $\KK$ is now restricted to the first
Brillouin zone and $\GG$ are the same reciprocal vectors of the
photonic lattice. The resulting equation in Fourier space is
\begin{widetext}
\begin{equation}
\sum_{\GG'}\left[\frac{\hbar^2|\kk+\GG|^2}{2M_{\mathrm{ex}}}
\,\delta_{\GG,\GG'}+V(\GG-\GG')\right]\,F(\KK+\GG')=
E_{\KK}\,F(\KK+\GG)
\,\label{eq:schrodinger2},
\end{equation}
\end{widetext}
where the Fourier matrix $V(\GG,\GG')\equiv V(\GG-\GG')$ has
expressions similar to those for $\epsilon(\GG,\GG')$ in the
photonic problem. Equation~(\ref{eq:schrodinger2}) is solved
numerically, yielding quantized center-of-mass levels in the
periodic potential. By this procedure, the exciton levels are
labelled by the same quantum numbers as the electromagnetic modes,
namely a Bloch vector $\bf{K}$ and a discrete index $\nu$. The
exciton energy is written as $E^{(\mathrm{ex})}_{\KK
\nu}=E_{\mathrm{ex}}+ E_{\mathbf{K}\nu}$, where $E_{\mathrm{ex}}$
is the bare QW exciton energy (which is given for the specific QW
parameters), while $E_{\mathbf{K}\nu}$ is the center-of-mass
quantization energy in the in-plane potential
$V(\mathbf{R}_{\parallel})$. It turns out that
$E_{\mathbf{K}\nu}\sim 10^{-2}-10^{-1}$~meV for typical PhC slab
patterns, depending on $a$ and $r/a$. Now we may introduce exciton
creation (destruction) operators for center-of-mass eigenmodes,
$\hat{b}_{\mathbf{K}\nu}^{\dagger}$ ($\hat{b}_{\mathbf{K}\nu}$).
In second quantization, the hamiltonian of the bare exciton field
is finally given by
\begin{equation}
\hat{H}_{\mathrm{ex}}= \sum_{\mathbf{K},\nu}E^{(\mathrm{ex})}_{\KK
\nu}\,\hat{b}_{\mathbf{K}\nu}^{\dagger} \hat{b}_{\mathbf{K}\nu}
\,\label{excham},
\end{equation}
where for $n_{\mathrm{ex}}\ll 1/a^2_{2D}$ the excitonic operators
obey commutation relations
\begin{equation}
[\hat{b}_{\mathbf{K}\nu},\hat{b}_{\mathbf{K}'
\nu'}^{\dagger}]\simeq
\delta_{\mathbf{K},\mathbf{K}'}\delta_{\nu,\nu'} \,,\,
[\hat{b}_{\mathbf{K}\nu},\hat{b}_{\mathbf{K}' \nu'}]=
[\hat{b}_{\mathbf{K}\nu}^{\dagger},\hat{b}_{\mathbf{K}'
\nu'}^{\dagger}]\simeq 0
\,\label{commrelexc}.
\end{equation}
Broadening of the exciton spectral lines can be taken into account
phenomenologically by introducing an imaginary part of mode
energies $\gamma_{\mathrm{ex}}$, taken as a $\KK$- and
$\nu$-independent parameter.

\subsection{Exciton-photon interaction hamiltonian}

The interaction between exciton and photon states must conserve
the Bloch vector, i.e., $\bf{K}=\kk$ (modulo a reciprocal lattice
vector). Thus, we will use the same notation for exciton and
photon wave vectors. However, in general a photonic mode with band
index $n$ couples to exciton center-of-mass levels with any $\nu$.
The interaction is determined by a matrix element of the full
hamiltonian, as first shown in
Refs.~\onlinecite{hopfield58,agranovich} for bulk
exciton-polaritons and later extended to quantum-confined
systems.\cite{savona94,pau95prb,savasta96,panzarini99} The
classical minimal coupling hamiltonian is given by
\begin{widetext}
\begin{equation}
\mathcal{H}_{\mathrm{I}}=-\frac{e}{2m_0c}\sum_{j=1}^{N}\left\{\mathbf{A}(\mathbf{r}_j)
\cdot\mathbf{p}_j+
\mathbf{p}_j\cdot\mathbf{A}(\mathbf{r}_j)\right\}+\frac{e^2}{2m_0c^2}\sum_{j=1}^{N}
|\mathbf{A}(\mathbf{r}_j)|^2
\,\label{hamintcl},
\end{equation}
\end{widetext}
where $m_0$ is the free electron mass, $\mathbf{r}_{j}$
($\mathbf{p}_{j}$) are the position (momentum) variables of the QW
electrons, and the sum runs over all the electrons in the system.
In Eq.~(\ref{hamintcl}), we have retained both
$\mathbf{A}\cdot\mathbf{p}$ and $\mathbf{p}\cdot\mathbf{A}$ terms,
because the generalized Coulomb gauge~(\ref{eq:gauge}) does not
allow in general for the commutation of these two operators.
Taking into account the presence of a non-local potential, we can
write this hamiltonian in second-quantized form as
$\hat{H}_{\mathrm{I}}=\hat{H}_{\mathrm{I}}^{(1)}+\hat{H}_{\mathrm{I}}^{(2)}$,
where the two contributions are\cite{girlanda81,note_commutator}
\begin{eqnarray}
\hat{H}_{\mathrm{I}}^{(1)}&=&i\frac{e}{\hbar
c}\sum_{j=1}^{N}\hat{\mathbf{A}}(\mathbf{r}_j)
\cdot[\hat{\mathbf{r}}_j,\hat{H}_{\mathrm{ex}}]
\,\label{hamintcl1},\\
\hat{H}_{\mathrm{I}}^{(2)} &=& -i\frac{e^2}{2\hbar
c^2}\sum_{j=1}^{N}
[\hat{\mathbf{A}}(\mathbf{r}_j)\cdot\hat{\mathbf{r}}_j,
\hat{\mathbf{A}}(\mathbf{r}_j)\cdot\hat{\mathbf{v}}_j]
\,\label{hamintcl2},
\end{eqnarray}
and the operator
$\hat{\mathbf{v}}_j=\hat{\mathbf{p}}_j/m_0=(i\hbar)^{-1}
[\hat{\mathbf{r}}_j,\hat{H}_{\mathrm{ex}}]$
directly follows from Heisenberg equation of motion for
$\hat{\mathbf{r}}_j$. Introducing exciton operators and resolving
the commutator, one can write, e.g., for the first term
\begin{equation}
\hat{H}_{\mathrm{I}}^{(1)}=-i\frac{e}{\hbar c}\sum_{\kk,\nu}
E^{(\mathrm{ex})}_{\kk \nu}\langle\Psi^{(\mathrm{ex})}_{\kk
\nu}|\sum_{j}\hat{\mathbf{A}}(\mathbf{r}_{j})\cdot
\mathbf{r}_{j}|0\rangle \, \hat{b}_{\kk
\nu}^{\dagger}+\mathrm{h.c.} \, \label{hamintsq},
\end{equation}
where $\Psi^{(\mathrm{ex})}_{\kk \nu}$ is the many-body exciton
wavefunction while $|0\rangle $ is the crystal ground state.
Expanding the vector potential as in Eq.~(\ref{sqAfield}), and
expressing also $\hat{H}_{\mathrm{I}}^{(2)}$ in terms of exciton
operators, after some manipulation the two terms of the
second-quantized interaction hamiltonian are obtained in the form
\begin{eqnarray}
&&\hat{H}_{\mathrm{I}}^{(1)} =  i\sum_{\kk, n, \nu} C_{\kk n \nu}
                 (\hat{a}_{\kk n}+\hat{a}_{-\kk n}^{\dagger})
                 (\hat{b}_{-\kk \nu}-\hat{b}_{\kk \nu}^{\dagger}),\label{eq:qwham}\\
&&\hat{H}_{\mathrm{I}}^{(2)} = \sum_{\kk,\nu, n, n'}D_{\kk \nu n
n'}
       (\hat{a}_{-\kk n}+\hat{a}_{\kk n}^{\dagger})
       (\hat{a}_{\kk n'}+\hat{a}_{-\kk n'}^{\dagger})
       \,\label{eq:qwham_bis}.
\end{eqnarray}
The coupling matrix element $C_{\kk n \nu}$ between exciton and
photon states at a given $\kk$ is calculated as\cite{note_phase}
\begin{equation}
C_{\kk n \nu } =  E^{(\mathrm{ex})}_{\kk \nu}\left(\frac{2\pi
e^2\hbar\omega_{\kk n}} {\hbar^2c^2}\right)^{1/2}
\langle\Psi^{(\mathrm{ex})}_{\kk \nu}|\sum_{j}\mathbf{A}_{\kk
n}(\mathbf{r}_{j})\cdot \mathbf{r}_{j}|0\rangle
\,\label{eq:coupling} ,
\end{equation}
while $D_{\kk \nu n n'}={C_{\kk n \nu}^{\ast}C_{\kk n'
\nu}}/{E^{(\mathrm{ex})}_{\kk \nu}}$.

The integral in Eq.~(\ref{eq:coupling}) can be expressed in terms
the oscillator strength $f$ of the excitonic transition, which is
generally defined as\cite{andreani_erice}
\begin{equation}
f_{\hat{\mathbf{e}}}=\frac{2m_0\Omega_{\mathrm{ex}}}{\hbar}
|\langle\Psi^{(\mathrm{ex})}_{\kk
\nu}|\hat{\mathbf{e}}\cdot\sum_{j} \mathbf{r}_{j}|0\rangle|^2 \,
\label{eq:oscstrength},
\end{equation}
where $\Omega_{\mathrm{ex}}=E^{(\mathrm{ex})}_{\kk \nu}/\hbar$ and
$\hat{\mathbf{e}}$ is the polarization unit vector of the exciton.
For a QW exciton the oscillator strength per unit area is
calculated as
\begin{equation}
\frac{f_{\hat{\mathbf{e}}}}{S}=\frac{2m_0\Omega_{\mathrm{ex}}}{\hbar}
|\hat{\mathbf{e}}\cdot\mathbf{r}_{\mathrm{c}\mathrm{v}}|^2
\,\left|\int f({\mbox{\boldmath$\rho$}}=0,z,z)\mathrm{d}z\right|^2
\, \label{oscstrengthqw},
\end{equation}
in which $\mathbf{r}_{\mathrm{c}\mathrm{v}}=\langle
u_{\mathrm{c}0}| \mathbf{r}|u_{\mathrm{v}0}\rangle$ is the dipole
matrix element between the single-particle Bloch functions in the
valence and conduction bands of the bulk crystal. We are
considering the ground-state heavy-hole (HH) exciton, whose
optically active states are polarized in the $xy$ plane and are
doubly degenerate. Thus, HH QW excitons preferentially couple to
TE-like modes in the photonic structure. Expressing the exciton
wave function in terms of the QW envelope function
(\ref{envelopeqw}), it is easy to see that the coupling energy
(\ref{eq:coupling}) depends on the oscillator strength per unit
area, Eq.~(\ref{oscstrengthqw}), as well as on the spatial overlap
between the exciton center-of-mass wavefunction and the photonic
mode profile in the QW plane. Assuming the mode profile to be
uniform along the QW thickness, we can express the coupling matrix
element as\cite{note_phase}
\begin{equation}
C_{\kk n \nu } \simeq -i\left(\frac{\pi\hbar^2e^2}
{m_0}\frac{f}{S}\right)^{1/2}\, \int \hat{\mathbf{e}} \cdot
\mathbf{E}_{\kk n}({\mbox{\boldmath$\rho$}},z_{\mathrm{QW}})
F_{\kk \nu}^{\ast}({\mbox{\boldmath$\rho$}}) \,
\mathrm{d}{\mbox{\boldmath$\rho$}} \, \label{eq:coupling2} ,
\end{equation}
where we approximated $E^{(\mathrm{ex})}_{\kk \nu}\simeq
\hbar\omega_{\kk n}$, assuming close-to-resonance
coupling.\cite{note_coupling} We notice that for MQWs the coupling
energy (\ref{eq:coupling2}) can be multiplied by
$\sqrt{N_{\mathrm{QW}}}$, where $N_{\mathrm{QW}}$ is the effective
number of QWs coupled to the photonic mode. Within our GME
formalism, the integral in Eq.~(\ref{eq:coupling2}) can be
straightforwardly calculated in Fourier space and we finally get
\begin{equation}
C_{\kk n \nu } \simeq -i\left(\frac{\pi \hbar^2e^2}
{m_0}\frac{f}{S}\right)^{1/2}\, \sum_{\GG} \hat{\mathbf{e}} \cdot
\mathbf{E}_{\kk+\GG, n}(z_{\mathrm{QW}}) F_{\kk+\GG,\nu}^{\ast} \,
\label{eq:integral},
\end{equation}
where $\mathbf{E}_{\kk+\GG, n}(z_{\mathrm{QW}})$ is the Fourier
transform of the mode electric field at the QW vertical
position.\cite{note_overlap} Thus, all parameters of the
interaction hamiltonian are obtained in terms of the electric
field coefficients calculated by the GME method and of the exciton
coefficients corresponding to quantized center-of-mass levels.

\subsection{Diagonalization of the total hamiltonian}

The full quantum hamiltonian describing the coupled QW exciton and
PhC slab modes is given by
\begin{equation}
\hat{H}_{\mathrm{tot}} =
\hat{H}_{\mathrm{ph}}+\hat{H}_{\mathrm{ex}}+
\hat{H}_{\mathrm{I}}^{(1)}+\hat{H}_{\mathrm{I}}^{(2)}\,\label{totham},
\end{equation}
where $\hat{H}_{\mathrm{ph}}$, $\hat{H}_{\mathrm{ex}}$ and the two
interaction hamiltonians are explicitly given in
Eqs.~(\ref{photham}), (\ref{excham}), and
(\ref{eq:qwham})-(\ref{eq:qwham_bis}). In the bare photonic
dispersion, we do not consider the zero-point energy term. Not
surprisingly, Eq.~(\ref{totham}) has a formal analogy with the
second-quantized exciton-photon hamiltonian in
bulk,\cite{hopfield58} planar MCs,\cite{savona94} or pillar
MCs.\cite{panzarini99} Such hamiltonian is valid in the linear
regime, i.e., under low excitation. Non-linear terms including
exciton-exciton scattering and saturation of exciton-photon
coupling\cite{ciuti00,ciuti03sst} are not considered here.

The total hamiltonian of the exciton-photon coupled system is
diagonalized by using a generalized Hopfield
transformation.\cite{hopfield58,savona94,savasta96,panzarini99}
New destruction (creation) operators $\hat{p}_{\kk}$
($\hat{p}_{\kk}^{\dagger}$) are defined as
\begin{equation}
\hat{p}_{\kk}=\sum_{n}w_{\kk n}a_{\kk n}+\sum_{\nu}x_{\kk
\nu}b_{\kk \nu}+ \sum_{n}y_{\kk n}a_{-\kk
n}^{\dagger}+\sum_{\nu}z_{\kk \nu}b_{-\kk
\nu}^{\dagger}\,\label{poloperator},
\end{equation}
which still satisfy Bose commutation relations
\begin{equation}
[\hat{p}_{\kk},\hat{p}_{\kk'}^{\dagger}]=
\delta_{\kk,\kk'}\,,\,\,\, [\hat{p}_{\kk},\hat{p}_{\kk'}]=
[\hat{p}_{\kk}^{\dagger},\hat{p}_{\kk'}^{\dagger}]=0 \,
\label{commrelp}.
\end{equation}
The condition for the total hamiltonian to be diagonal in terms of
$\hat{p}_{\kk}$, $\hat{p}_{\kk'}^{\dagger}$ is
\begin{equation}
[\hat{p}_{\kk},\hat{H}_{\mathrm{tot}}]=
\hbar\Omega_{\kk}\hat{p}_{\kk}\,\label{condition}.
\end{equation}
The transformation, which leads to a non-hermitian eigenvalue
problem, applies also to a hamiltonian that includes dissipative
terms. This is the present case when the imaginary part of the
frequency for quasi-guided photonic modes, as well as the exciton
linewidth arising from non-radiative processes, are included in
the terms $\hat{H}_{\mathrm{ph}}$, $\hat{H}_{\mathrm{ex}}$ of
Eq.~(\ref{totham}). The eigenvalue problem can be written in the
form
\begin{equation}
\mathbf{\mathbb{M}}_{\kk}\vec{\mathbf{v}}_{\kk}=
\hbar\Omega_{\kk}\vec{\mathbf{v}}_{\kk}\, \label{hopfield},
\end{equation}
where $\mathbf{\mathbb{M}}_{\kk}$ is the generalized Hopfield
matrix at a specific wave vector $\kk$, and
$\vec{\mathbf{v}}_{\kk}$ is a generalized vector constituted by
the expansion coefficients of Eq.~(\ref{poloperator}) (see
Appendix). Diagonalization of Eq.~(\ref{hopfield}) gives directly
the complex eigenenergies $\hbar\Omega_{\kk}$ corresponding to
mixed excitations of radiation and matter. Depending on the
interplay between exciton-photon coupling and their respective
losses, the system can be either in a weak or in a strong-coupling
regime. In the latter case, $\hbar\Omega_{\kk}$ calculated for any
$\kk$ in the first BZ gives rise to the full spectrum of
\textit{photonic crystal polaritons}.

\begin{figure}[t]
\begin{center}
\includegraphics[width=0.45\textwidth]{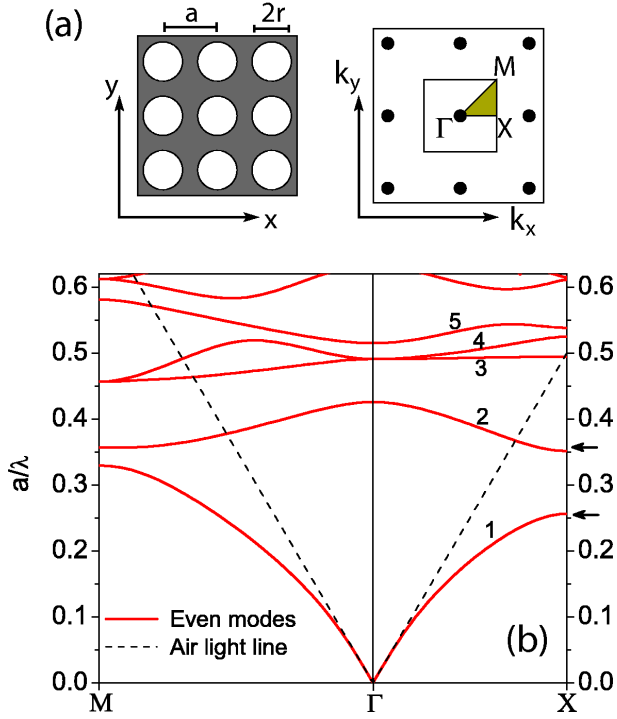}
\caption{(color online) (a) Square lattice of air holes: direct
and reciprocal lattices with Brillouin zone and main symmetry
points, (b) photonic mode dispersion (even modes,
$\sigma_{xy}=+1$) when the lattice is embedded in a high-index
($n_{\mathrm{diel}}=3.43$) photonic crystal membrane. Parameters
are: $r/a=0.34$, $d/a=0.3$. The first few modes are labelled by a
band number. Dashed lines represent the light dispersion in
air.\label{fig2} }
\end{center}
\end{figure}

\section{Numerical results: the square lattice PhC slab}\label{results}

In this Section we apply the present theory to a square lattice of
air holes in a symmetric PhC slab with air claddings. We consider
a GaAs membrane with a dielectric constant
$\epsilon_{\mathrm{diel}}=12.46$, typical of this material in the
near infrared. A schematic representation of the direct and
reciprocal lattices is shown in Fig.~\ref{fig2}(a). The photonic
lattice is characterized by the inter-hole separation, $a$, and
the hole radius, $r$. In reciprocal space, the BZ is evidenced and
the main symmetry points $\Gamma$, X, and M are defined. In
Fig.~\ref{fig2}(b) we show a typical photonic band diagram
calculated by using the GME method,\cite{andreani06prb} for a PhC
membrane with thickness $d/a=0.3$ and hole radius $r/a=0.34$. The
dispersion is shown in dimensionless frequency units [$\omega
a/(2\pi c)=a/\lambda$], and only for modes with even parity
($\sigma_{xy}=+1$) with respect to a horizontal mirror plane at
the middle of the slab. Such modes have mainly in-plane polarized
electric field (also defined TE-like
modes),\cite{johnson_book,andreani06prb} and they are dominantly
coupled to HH exciton states in the QW. No higher-order modes of
the slab waveguide are present for this structure in the frequency
range of Fig.~\ref{fig2}(b). The first five photonic bands are
labelled with integer numbers. Mode 1 and part of mode 2 lie below
the air light line and are therefore truly guided. In principle,
such modes have infinite lifetime, and can leak radiation out of
the slab plane only due to disorder in the photonic
structure,\cite{gerace04ol,gerace05pn} which we neglect in the
present paper. On the contrary, the modes are quasi-guided when
their dispersion falls above the light line in the first BZ. These
modes are coupled to the continuum of radiative PhC modes at the
same energy. Physically, quasi-guided modes are lossy due to
out-of-plane diffraction, thus acquiring a finite radiative
linewidth $2\gamma_{\kk n}$ as discussed in
Subsec.~\ref{subsec:emfield} and calculated by the GME method.

\begin{figure}[t]
\begin{center}
\includegraphics[width=0.45\textwidth]{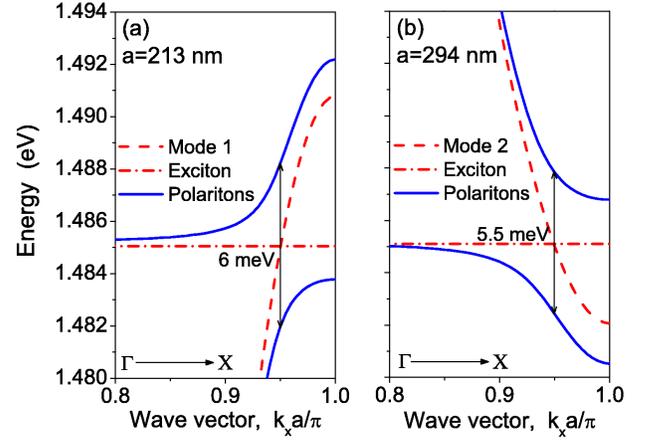}
\caption{(color online) Guided polariton dispersion for a photonic
mode interacting with a QW exciton at $E_{\mathrm{ex}}=1.485$ eV,
close to the BZ edge (X-point); the uncoupled mode dispersions are
shown with dashed and dot-dashed lines, respectively. Parameters
of the photonic structure are: $r/a=0.34$, $d/a=0.3$. Panel (a):
coupling to the first photonic mode, lattice constant $a=213$ nm.
Panel (b): coupling to the second mode, $a=294$ nm. \label{fig3}}
\end{center}
\end{figure}

\begin{figure}[b]
\begin{center}
\includegraphics[width=0.45\textwidth]{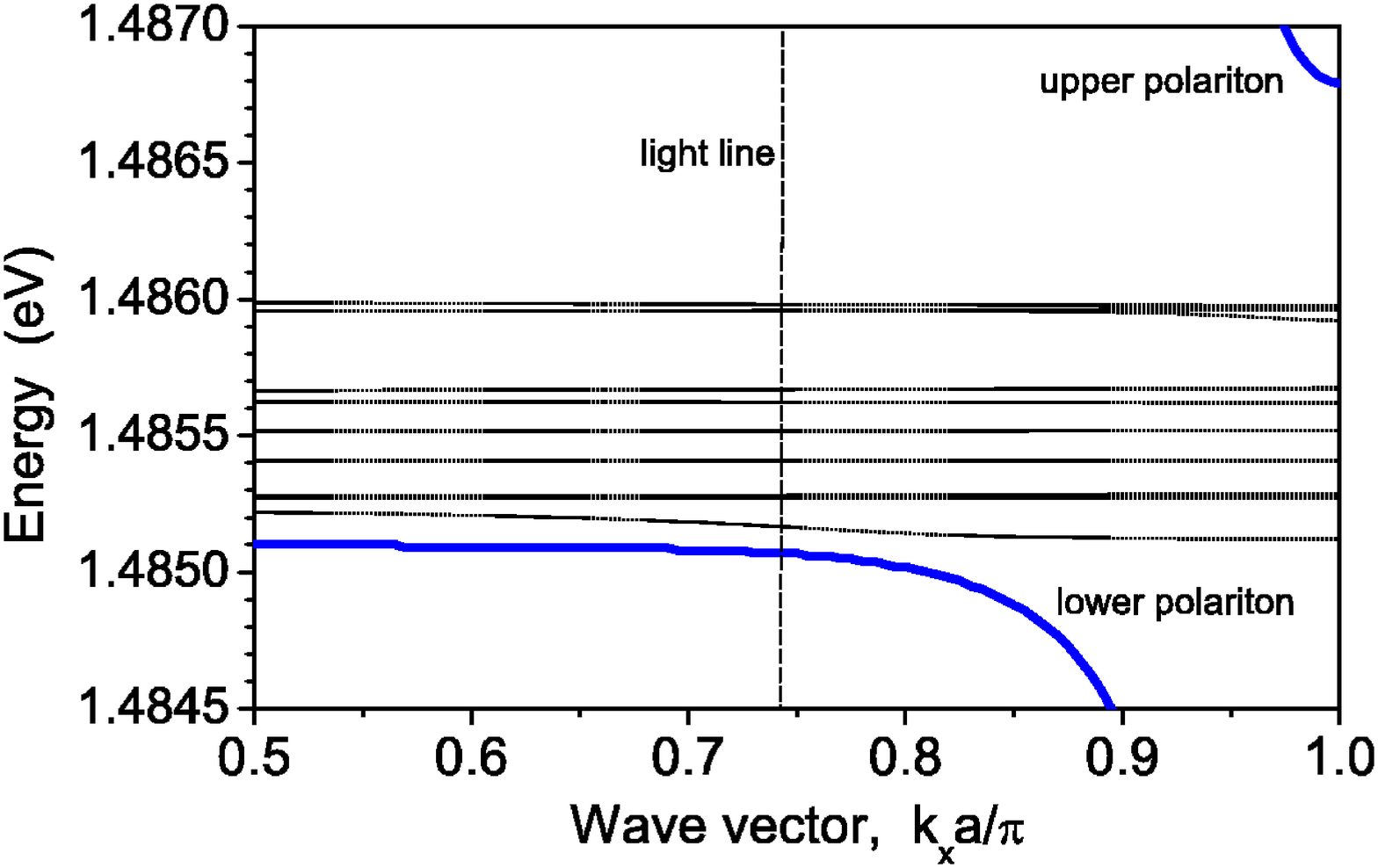}
\caption{(color online) Close-up of Fig. \ref{fig3}(b) with
exciton center-of-mass quantized levels explicitly
shown.\label{fig4} }
\end{center}
\end{figure}

We consider a single QW grown exactly in the middle of the GaAs
layer, which maximizes coupling to the electric field for
fundamental slab modes with even parity. The oscillator strength per
unit area will be assumed to be $f/S=4.2\times 10^{12}$ cm$^{-2}$,
typical of the HH exciton in an In$_x$Ga$_{1-x}$As QW with thickness
$L_{\mathrm{QW}}=8$~nm and low In
content.\cite{weisbuch92,houdre94a,iotti97} The exciton energy is
taken to be $E_{\mathrm{ex}}=1.485$ eV and the total exciton mass
$M_{\mathrm{ex}}=0.18 m_0$.\cite{andreani_erice} For the exciton
linewidth we assume a value $2\gamma_{\mathrm{ex}}=1$ meV.  Very
high-quality QW structures can be currently grown to achieve
linewidths smaller than 0.5~meV. Here we stay on the conservative
side, taking into account possible broadening induced by the
presence of interface defects at the hole boundaries. These
considerations are also supported by preliminary experimental data
on square lattice PhC slabs with embedded QWs, which show no
sizeable increase of the exciton linewidth as compared to an
unpatterned area of the same sample (for an initial linewidth on the
order of 2 meV).\cite{unpublished} With the parameters given here,
the exciton-photon coupling matrix element calculated from
Eq.~(\ref{eq:coupling2}) is on the order of a few meV, with slight
variations depending on the photonic mode of interest, and the
occurrence of strong or weak coupling for quasi-guided modes is
governed by the photonic mode linewidth rather than by the exciton
linewidth.\cite{note_parameters}

\begin{figure*}[!ht]
\begin{center}
\includegraphics[width=0.9\textwidth]{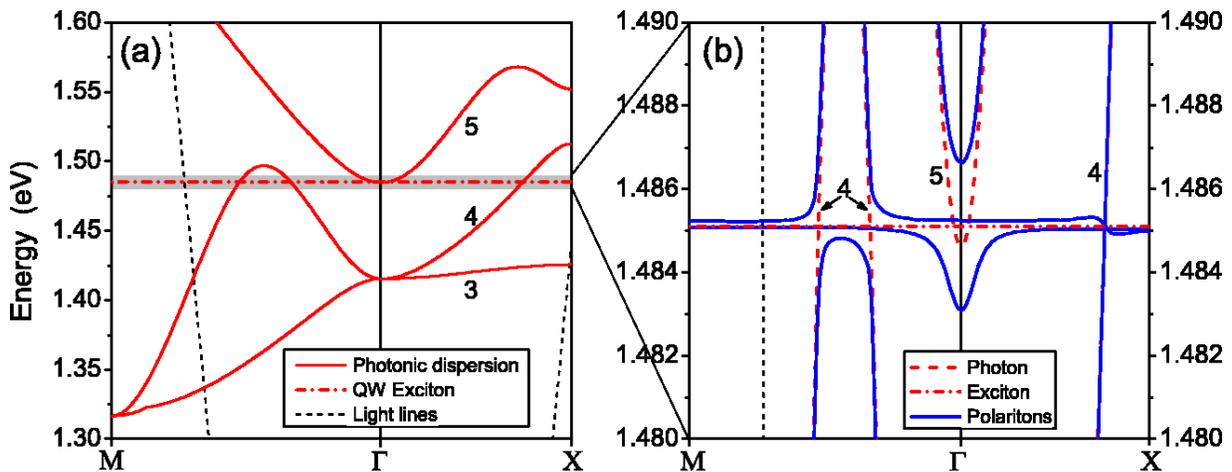}
\caption{(color online) (a) Photonic mode dispersion around the QW
exciton energy, $E_{\mathrm{ex}}=1.485$ eV, for structure
parameters: $r/a=0.34$, $d/a=0.3$, $a=430$ nm. (b) Close up for
the 10 meV energy range of interest, showing the solution for the
coupled exciton-photon system (full line) together with the bare
exciton and photon dispersions (dashed).\label{fig5} }
\end{center}
\end{figure*}

\subsection{Guided Polaritons}\label{subsec:guided}

Due to the scalability of Maxwell equations, the PhC lattice can be
engineered in order to have resonance between a given QW exciton
energy and the desired photonic mode at any specific point in the
BZ. When the exciton-photon resonance occurs below the light line
(and the exciton linewidth is sufficiently small), the system is
always in the strong coupling regime. The new quasi-particles that
describe the eigenstates of the system are guided PhC polaritons,
i.e., their photonic part is trapped within the high-index slab
through total internal reflection.

In Fig.~\ref{fig3}(a) and (b) we consider the interaction of a QW
exciton with photonic modes 1 and 2 from the band dispersion shown
in Fig.~\ref{fig2}(b), respectively. The photonic lattice is
engineered to have the resonance condition close to the BZ edge
along the $\Gamma$X direction. In one case, the lattice constant
is set to the value $a=213$ nm, and in the other we assume $a=294$
nm, so that the resonance at $E=1.485$ eV is for $k_x\simeq
0.95\pi/a$ [see also arrows in Fig.~\ref{fig2}(b)]. As it can be
seen from Fig.~\ref{fig3}, the dispersion of bare exciton and
photon modes is strongly modified in both cases, giving rise to
sizable anticrossings. It should be noted that in Fig.~\ref{fig3},
as well as in the following figures of Secs.
\ref{subsec:radiative} and \ref{subsec:discussion}, the uncoupled
exciton center-of-mass levels are not explicitly shown. We rather
prefer to show the bare exciton and photon dispersions as compared
to the strongly coupled polaritonic dispersion. The calculated
vacuum Rabi splitting is $\hbar\Omega_{\mathrm{R}}=6$ meV for the
first mode and $\hbar\Omega_{\mathrm{R}}=5.5$ meV for the second
one, respectively. The exciton-photon coupling is dependent on the
specific band of interest, due to the different spatial profile of
the corresponding electric field and thus to the modified overlap
with the exciton center-of-mass wavefunctions. In any case, we
point out that such values obtained with a \textit{single} quantum
well are comparable to those commonly achieved for MC polaritons
with \textit{six} QWs.\cite{houdre94a} The physical reason is the
increased exciton-photon coupling of Eq.~(\ref{eq:coupling2}), due
to better confinement in the vertical direction of a high-index
dielectric slab compared to a MC with low-index contrast
distributed Bragg reflectors.

A close up of the dispersion diagram showing the center-of-mass
quantized levels (with $M_{\mathrm{max}}=10$) is given in
Fig.~\ref{fig4}, for parameters as in Fig.~\ref{fig3}(b). It can be
noticed that the quantization energy is $\lesssim$~0.1 meV in this
specific case. These levels represent the effect of spatial
dispersion, well known for bulk
exciton-polaritons,\cite{andreani_erice} in the presence of
center-of-mass quantization. Only the first few exciton levels are
strongly interacting with the photonic mode of interest, due to
symmetry of the corresponding envelope function and electric field.
We point out that such effects cannot be captured by theoretical
treatments based on a semiclassical solution of Maxwell's
equations,\cite{yablonskii01,shimada02ieee} in which the effect of
spatial dispersion is neglected. Besides giving directly the
dispersion of the mixed exciton-photon modes, the quantum theory
developed here is also the starting point for studying polariton
interactions and nonlinear processes, like in the case of bulk and
microcavity systems.

\subsection{Radiative Polaritons}\label{subsec:radiative}

When a QW exciton is resonant with a quasi-guided photonic mode,
weak or strong coupling regimes may occur depending on the
specific situation. We show in Fig.~\ref{fig5} the case of a PhC
slab of lattice constant $a=430$~nm, in which the exciton is
resonant with different photonic modes within the first BZ [namely
modes labelled with indices 4 and 5 in Fig.~\ref{fig2}(b)]. In
Fig.~\ref{fig5}(a), the bare photonic mode dispersion around
$E_{\mathrm{ex}}=1.485$ eV is shown. It is interesting to notice
the existence of a photonic band minimum for mode 5 at the
$\Gamma$ point: such feature leads to a quasi-particle dispersion
similar to MC polaritons, as it will be discussed. Actually, the
resonance condition occurs simultaneously with different modes
along $\Gamma$M and $\Gamma$X. In Fig.~\ref{fig5}(b), the
dispersion of the exciton-photon coupled modes is shown in a
restricted energy range around the exciton resonance. Notice that
there are five resonant points between QW excitons and PhC slab
modes in the dispersion diagram, leading to a variety of
situations for the coupled modes. Along $\Gamma$M, clear
anticrossings can be seen with photonic modes 4 and 5, which are
fingerprints of the strong coupling regime. In this case, the
intrinsic radiative linewidth of bare photonic states is lower
than the exciton-photon coupling energy. As the QW exciton is
resonant with mode 4 for two different wave vectors along
$\Gamma$M, we observe two anticrossings above the light line in
the middle of the BZ. This peculiar effect is due to the light
dispersion engineering allowed in PhC structures. Along $\Gamma$X,
resonance with mode 5 gives strong coupling, while resonance with
mode 4 at larger in-plane wave vector gives a crossing of the bare
excitonic and photonic dispersions, meaning that the system is in
weak coupling. We will analyze separately these distinct regimes
in the following.

\begin{figure}[t]
\begin{center}
\includegraphics[width=0.4\textwidth]{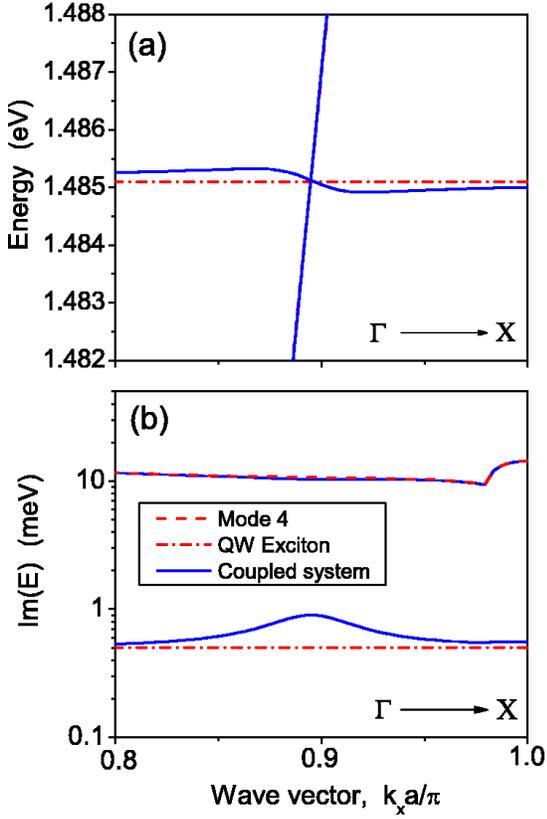}
\caption{(color online) (a) Real and (b) imaginary parts of
complex eigenenergies in the weak exciton-photon coupling regime
close to the BZ edge along the $\Gamma$X direction. The bare
exciton (dot-dashed) and photon (dashed) complex energies are also
shown (the real part of photon dispersion nearly coincides with
the coupled system solution). Structure parameters are as in
Fig.~\ref{fig5}. \label{fig6} }
\end{center}
\end{figure}

We show in Fig.~\ref{fig6} the complex dispersion of the
eigenmodes in the region of weak coupling, as compared to the bare
QW exciton and photon energies. As seen in Fig.~\ref{fig6}(a),
photonic mode 4 is in weak coupling with all quantized exciton
states (only one is shown here, for clarity) close to the X point.
A slight modification of the exciton dispersion occurs close to
resonance for the center-of-mass quantized mode having the same
symmetry as the resonant photonic mode. The photonic mode
dispersion is not perturbed at all by the presence of the QW.
Remarkably, the plot of the intrinsic imaginary part calculated
for the bare quasi-guided photonic mode in Fig.~\ref{fig6}(b),
evidences a linewidth close to $2\gamma_{\mathrm{ph}}\sim 20$ meV,
which explains the weak coupling regime for this particular case.
Furthermore, the imaginary part of the photonic mode energy
extracted from the solution of the Hopfield matrix shows no
modification. On the contrary, and quite interestingly, the
imaginary part of the QW exciton energy, initially set to
$\gamma_{\mathrm{ex}}=0.5$ meV for the unperturbed state, is
sensitively increased around the resonance. This indicates the
occurrence of an enhancement of spontaneous emission rate or
Purcell effect, which may be observed by time-resolved
experiments.

\begin{figure}[t]
\begin{center}
\includegraphics[width=0.4\textwidth]{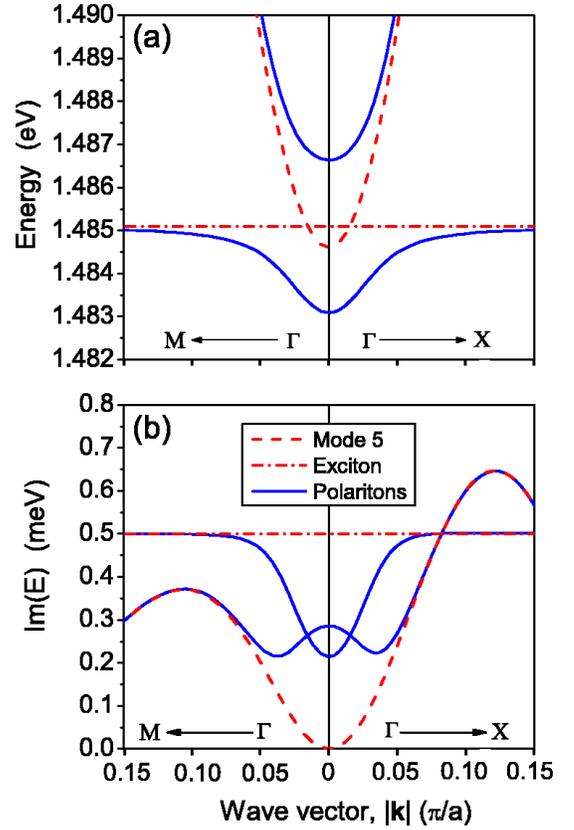}
\caption{(color online) (a) Real and (b) imaginary parts of
complex eigenenergies in the strong exciton-photon coupling regime
close to normal incidence ($\Gamma$-point of the BZ) along the two
main symmetry directions; the dispersion of uncoupled modes is
also shown. Structure parameters are as in
Fig.~\ref{fig5}.\label{fig7} }
\end{center}
\end{figure}

In Fig.~\ref{fig7} we analyze the strong coupling regime above the
light line, giving rise to radiative PhC polariton states. Mode 5
has a vanishing intrinsic linewidth at $\Gamma$ [see imaginary
part of the bare photonic mode in Fig.~\ref{fig7}(b), dashed
line], leading to a vacuum Rabi splitting. The real parts of the
mode energies are shown around $\kk=0$ along the $\Gamma$M and
$\Gamma$X symmetry directions. Looking at the bare QW exciton and
mode 5 dispersions, we notice that the resonance condition is not
exactly at $\Gamma$, but at small wave vectors ($|\kk|\simeq
0.02\pi/a$). The exact resonant wave vector can be also inferred
from Fig.~\ref{fig7}(b), corresponding to the upper and lower
polariton imaginary parts being coincident and equal to the
average of bare exciton and photon values. As the bare photonic
linewidth goes to zero at $\Gamma$, the polariton imaginary part
is about half of the bare QW exciton one at $\kk=0$. This
\textit{linewidth averaging} effect is well known for MC
polaritons\cite{weisbuch92,houdre94a} and is a fingerprint of the
occurrence of a vacuum Rabi splitting.

\subsection{Discussion and possible experimental verification}
\label{subsec:discussion}

\begin{figure}[t]
\begin{center}
\includegraphics[width=0.4\textwidth]{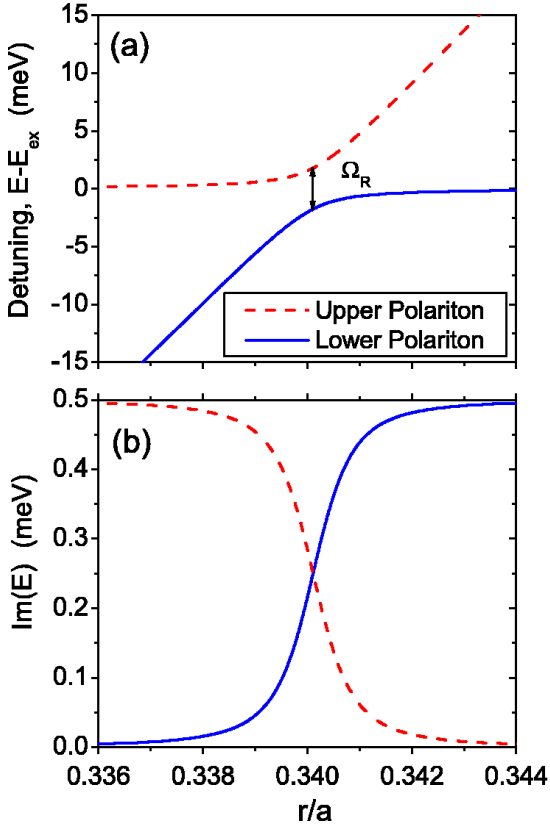}
\caption{(color online) (a) Real and (b) imaginary parts of upper
and lower polariton modes at $k=0$ ($\Gamma$-point) as a function
of hole radius. Structure parameters are as in Fig.~\ref{fig5}.
\label{fig8} }
\end{center}
\end{figure}

PhC slabs allow for a sensitive flexibility and tuning capability.
In particular, a \textit{lithographic tuning} (e.g., variation of
the lattice constant or the air hole radius) on different devices
patterned on the same chip is commonly used to ensure resonance of
a desired mode with an active medium. Furthermore, techniques such
as digital etching\cite{badolato05sci} have been recently employed
to achieve a fine tuning of photonic mode resonances. By using
these post-processing techniques, it would be possible to tune the
photonic mode of interest with respect to the QW exciton
resonance. In Fig.~\ref{fig8} we show a simulation of such a
procedure, in which the calculated detunings of the upper and
lower polariton branches from the bare exciton resonance at
$\kk=0$ are reported as a function of hole radius. The latter is
slightly increased from $r/a=0.336$ ($r\simeq 144.5$ nm) to
$r/a=0.344$ ($r\simeq 147.9$ nm). The typical anticrossing occurs
exactly on resonance in the real part of energy, and the
corresponding imaginary parts exchange each other from purely
excitonic to purely photonic and viceversa. Concerning the
significance of Fig.~\ref{fig8} with respect to state-of-the art
technological capabilities, wet chemical digital etching allows
for deposition of a few monolayers of surface oxide that can be
selectively removed, thus yielding a wavelength tuning of $\sim 1$
nm per etching cycle.\cite{kevin05apl} On the other hand, surface
AFM oxidation of the PhC membrane can lead, in principle, to
almost continuous tuning of the mode wavelength.\cite{kevin06apl}

Radiative PhC polaritons can be probed by angle-resolved
reflectance from the sample surface, as first done on 1D PhCs
filled with organic molecules.\cite{fujita98} The same kind of
experiment could be performed with semiconductor-based systems
discussed in this work. Indeed, the present quantum-mechanical
treatment of interacting photon and exciton states has been
compared with semiclassical calculations of the surface
reflectance, showing a very good agreement for the splitting in
strong coupling
regime.\cite{gerace04pss,andreani04pn,andreani05pss} For guided
polaritons, on the other hand, coupling to an external propagating
beam is prohibited due to the evanescent character of the
electromagnetic field in the claddings. In this case it would be
possible to perform angle-resolved attenuated total reflectance by
using a high-index prism. This technique was applied in
Refs.~\onlinecite{galli04prb,galli05prb} to probe the dispersion
of pure photonic states in line-defect PhC waveguides.

\begin{figure}[!ht]
\begin{center}
\includegraphics[width=0.4\textwidth]{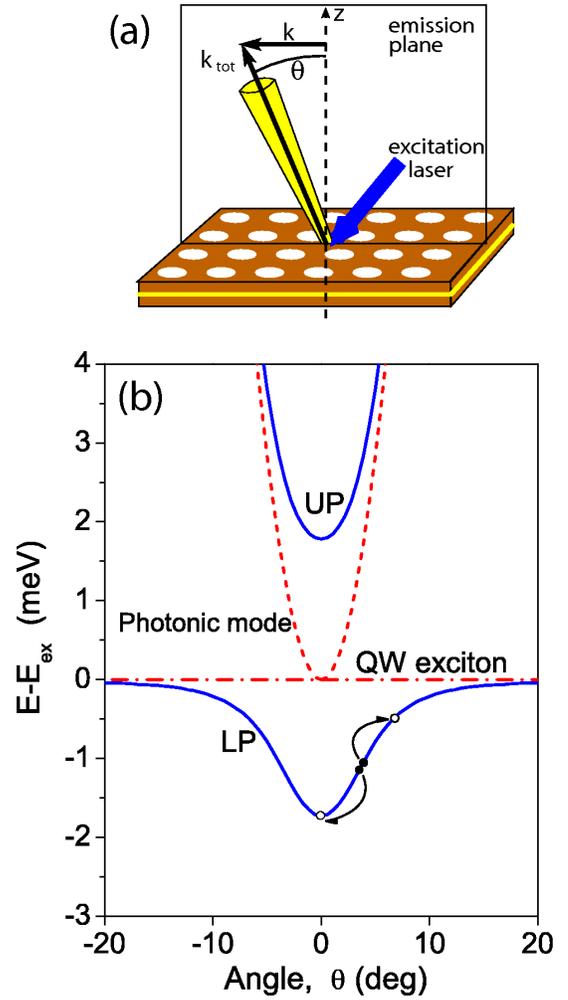}
\caption{(color online) (a) Schematic view of angular emission
from a PhC slab in a plane along a symmetry direction. The
emission angle, $\theta=\arcsin(kc/\omega)$, is defined. (b)
Calculated polariton angular dispersion in the $\Gamma$X
direction, at zero bare exciton-photon detuning (see
Fig.~\ref{fig8}); a schematic polariton-polariton scattering
process in the lower branch satisfying energy-momentum
conservation is also shown.\label{fig9} }
\end{center}
\end{figure}

We now focus on radiative properties close to $\kk=0$. As shown in
Fig.~\ref{fig7}, radiative polaritons can form with an energy
minimum at the $\Gamma$ point, due to the peculiar quasi-guided
photonic mode dispersion in a square lattice PhC membrane. The new
eigenmodes of the system can be excited by coherent or incoherent
pumping, and probed by emission or reflection at a fixed angle
$\theta$, which is defined in a vertical plane containing one of
the high symmetry directions of the PhC lattice. A schematic
picture of such an experimental configuration is shown in
Fig.~\ref{fig9}(a). The angle $\theta$ such that
$k=(\omega/c)\sin\theta$, due to conservation of in-plane
momentum. We consider here the angular dispersion along the
$\Gamma$X direction for the case of zero-detuned (at normal
incidence) bare exciton and photon modes. The calculated angular
dispersion is shown in Fig.~\ref{fig9}(b), in which we plot the
energies of the upper and lower polariton branches detuned from
the bare QW exciton energy (dot-dashed line).
It is interesting to notice the similarities between the angular
dispersion of Fig.~\ref{fig9}(b) and the usual diagram used to
illustrate the formation of a polariton trap in the energy minimum
at $\kk=0$.\cite{ciuti03sst} The peculiar properties of such an
angular dispersion have been used in the last few years to achieve
a number of outstanding results requiring non-linear parametric
processes of MC
polaritons.\cite{savvidis00,saba01nat,kasprzak06nat} In
particular, coherent population of the lowest energy polariton
state at $\kk=0$ can be realized by polariton-polariton scattering
from a precise point in the dispersion, which allows for
simultaneous energy and momentum conservation of the scattered
quasi-particles. Similar experiments could be realized also with
radiative PhC polaritons and a parametric process is schematically
illustrated in the lower polariton branch of Fig.~\ref{fig9}(b).
An extension of the present theory to include nonlinear terms
(such as exciton-exciton scattering)\cite{ciuti00} in the
hamiltonian represents a natural extension of this work. Moreover,
the recent observation of the long-sought Bose-Einstein
condensation of MC polaritons\cite{imamoglu96pra,porras02prb} by
use of II-VI materials\cite{kasprzak06nat} has enhanced interest
in low-dimensional polariton physics. The analogous effect in
III-V materials, like InGaAs/GaAs, has not been observed at time
of writing. A new route has been suggested to this
end,\cite{savona06pss} requiring efficient 0D confinement of MC
polaritons through the confinement of their photonic
part.\cite{eldaif06apl} This is strong motivation for experimental
as well as theoretical study of \textit{PhC cavity polaritons}, in
which ultra-high quality-factor and small mode volumes can be
achieved and enhance the radiation-matter coupling in an
unprecedented way.

\section{Conclusions}\label{concl}

In conclusion, we have presented a quantum theory to describe
radiation-matter coupling for quantum wells embedded in a
high-index photonic crystal slab with a generic pattern. The
numerical solution of classical Maxwell equations, on which the
present theory relies, has been previously reported within a
guided-mode expansion approach\cite{andreani06prb} which yields
not only the photonic mode energies but also the crucial photonic
linewidth parameter. After obtaining the second-quantized total
hamiltonian of the system in the linear regime of low excitation
density, we have diagonalized it through a generalized Hopfield
method. Thus, the complex eigenmodes of the exciton-photon coupled
system have been obtained. Both weak and strong coupling regimes
are treated within the present theory, allowing for an efficient
design of photonic lattices for specific purposes.

We have shown results on the square lattice PhC slab demonstrating
the formation of guided and radiative photonic crystal polaritons.
The latter are formed when the exciton-photon coupling exceeds the
intrinsic photon linewidth. In strong coupling, the vacuum Rabi
splitting at resonance is larger than for the corresponding
microcavity polaritons, due to the increased field confinement and
better overlap of exciton and photon wave fields in a PhC slab. As a
consequence, a more robust polariton effect can be envisioned in
such structures, which could be useful for nonlinear polariton
applications. Related to the latter point, we reported on a specific
lattice design for which the radiative polariton dispersion has a
minimum around normal incidence. Such dispersion closely mimics the
one obtained in usual microcavities, which acts as a polariton trap
and is at the origin of current research on nonlinear parametric
processes with exciton-polaritons. We believe that the present
results will stimulate further research on non-linear polariton
effects in semiconductor-based photonic crystals, which in turn
could connect with polariton quantum
optics\cite{ciuti04prb,edamatsu04,savasta05prl} as an emerging field
of research.

\appendix

\section{Generalized Hopfield method}\label{app:hopfield}

The Hopfield method to diagonalize the coupled exciton-photon
hamiltonian is equivalent to performing a Bogoljubov
transformation on bare exciton and photon operators. It was
originally used by Hopfield for the case of bulk
polaritons,\cite{hopfield58} and then generalized to polaritons in
planar MCs\cite{savona94,savasta96} and in
micro-pillars.\cite{panzarini99} Here we provide a detailed
derivation of the generalized Hopfield matrix for the case of PhC
slabs, which has to be diagonalized numerically. The
second-quantized total hamiltonian, Eq.~(\ref{totham}), and the
polariton operator expansion, Eq.~(\ref{poloperator}), are
substituted in Eq.~(\ref{condition}). The general commutation
relations for bosonic operators
\begin{eqnarray}
\left[\hat{a}_{\kk n},\hat{a}_{\kk' n'}^{\dagger} \hat{a}_{\kk'
n'}\right] &=& \delta_{\kk,\kk'} \delta_{n,n'}
\hat{a}_{\kk n} \nonumber\\
\left[\hat{a}_{\kk n}^{\dagger},\hat{a}_{\kk' n'}^{\dagger}
\hat{a}_{\kk' n'}\right] &=& - \delta_{\kk,\kk'} \delta_{n,n'}
\hat{a}_{\kk n}^{\dagger} \nonumber\\
\left[\hat{a}_{\kk n},\hat{a}_{\kk' n'} \hat{a}_{\kk'
n'}^{\dagger}\right] &=& \delta_{\kk,\kk'} \delta_{n,n'}
\hat{a}_{\kk n} \nonumber\\
\left[\hat{a}_{\kk n}^{\dagger},\hat{a}_{\kk' n'} \hat{a}_{\kk'
n'}^{\dagger}\right] &=& - \delta_{\kk,\kk'} \delta_{n,n'}
\hat{a}_{\kk n}^{\dagger} \,\label{app:commuation}
\end{eqnarray}
must be satisfied. Thus, after factorizing terms with common
operators, the following system of linear equations in the
variables $w_{n}$, $y_{n}$, $x_{\nu}$, and $z_{\nu}$ can be
derived (we drop the fixed subscript $\kk$ resp. $-\kk$, for
easier notation):
\begin{widetext}
\begin{eqnarray}
\hbar\omega_{n}\,w_{n}+\sum_{n',\nu}2D_{\nu \,n n'}\,w_{n'}-
\sum_{n',\nu}2D_{\nu \,n n'}\,y_{n'}-\sum_{\nu}iC_{n
\nu}\,x_{\nu}-\sum_{\nu}
iC_{n \nu}\,z_{\nu}&=&\hbar\Omega \, w_{n}\nonumber\\
+\sum_{n}iC_{n \nu}\,w_{n}-\sum_{n}iC_{n \nu}\,y_{n}+
E^{(\mathrm{ex})}_{\nu}\,x_{\nu}&=&\hbar\Omega\,x_{n}\nonumber\\
\sum_{n',\nu}2D_{\nu \,n n'}\,w_{n'}- \hbar\omega_{n}\,y_{n}-
\sum_{n',\nu}2D_{\nu\,n n'}\,y_{n'}-\sum_{\nu}iC_{n
\nu}\,x_{\nu}-\sum_{\nu}
iC_{n \nu}\,z_{\nu}&=&\hbar\Omega \, y_{n}\nonumber\\
-\sum_{n}iC_{n \nu}\,w_{n}+\sum_{n}iC_{n \nu}\,y_{n}-
E^{(\mathrm{ex})}_{\nu}\,z_{\nu}&=&\hbar\Omega\, z_{n}
\,\label{app:system}.
\end{eqnarray}
\end{widetext}
The sums in Eq.~(\ref{app:system}) must be truncated in order to
deal with finite matrices. If $N_{\mathrm{max}}$ photonic bands
and $M_{\mathrm{max}}$ excitonic levels at fixed $\kk$ are
retained in the expansion (\ref{poloperator}), the problem is
reduced to an eigenvalue equation
\begin{equation}
\mathbf{\mathbb{M}}\vec{\mathbf{v}}= \hbar\Omega\vec{\mathbf{v}}\,
\label{app:eigenv},
\end{equation}
where the matrix $\mathbf{\mathbb{M}}$ has dimension
$2(N_{\mathrm{max}}+M_{\mathrm{max}})\times
2(N_{\mathrm{max}}+M_{\mathrm{max}})$, and it is a generalization
of the $4\times 4$ matrix derived by Hopfield.\cite{hopfield58}
The vector $\vec{\mathbf{v}}$ is simply
\begin{equation}
\vec{\mathbf{v}}=(w_n,x_{\nu},y_n,z_{\nu})^{\mathrm{T}}\,
\label{app:vector},
\end{equation}
with $n=1,\ldots,N_{\mathrm{max}}$,
$\nu=1,\ldots,M_{\mathrm{max}}$. The explicit form of the
generalized Hopfield matrix is given by
\begin{equation}
\mathbf{\mathbb{M}}=\left(\begin{array}{c c c c}
{\mbox{\boldmath$\omega$}}+2\mathbf{D} & -i\mathbf{C} &
-2\mathbf{D} & -i\mathbf{C} \\
+i\mathbf{C} & \mathbf{E}
& -i\mathbf{C} & \mathbf{0} \\
+2\mathbf{D} & -i\mathbf{C} &
-{\mbox{\boldmath$\omega$}}-2\mathbf{D} & -i\mathbf{C} \\
-i\mathbf{C} & \mathbf{0} & +i\mathbf{C} & -\mathbf{E}
\end{array}\right) \, \label{app:hopfield_mtx}.
\end{equation}
The single blocks are respectively given by the
$N_{\mathrm{max}}\times N_{\mathrm{max}}$ matrices
\begin{equation}
{\mbox{\boldmath$\omega$}}+2\mathbf{D} =
\mat{\hbar\omega_{n}\delta_{n,n'}+\sum_{\nu}2D_{\nu \,n\,n'}} \,
\label{app:block1},
\end{equation}
by the $M_{\mathrm{max}}\times M_{\mathrm{max}}$ diagonal matrices
\begin{equation}
\mathbf{E}=\mat{E^{(\mathrm{ex})}_{\nu}\delta_{\nu,\nu'}}\,
\label{app:block2},
\end{equation}
and by the $N_{\mathrm{max}}\times M_{\mathrm{max}}$ matrices
\begin{equation}
\mathbf{C}=\mat{C_{n\,\nu}}\, \label{app:block3},
\end{equation}
while $\mathbf{0}$ denotes the $M_{\mathrm{max}}\times
M_{\mathrm{max}}$ matrix with zero entries. It should be noted
that both positive- and negative-defined energy values are present
in the Hopfield matrix. After the numerical diagonalization,
yielding the eigenvalues $\hbar\Omega_p=E_p+i\gamma_p$, with
$p=1,2,\ldots,2(N_{\mathrm{max}}+M_{\mathrm{max}})$, only those
with positive real part are eventually retained.

\begin{acknowledgments}
The authors are indebted with M.~Agio for collaboration and for
participating in the early comparisons between the present quantum
formalism and calculations based on a semiclassical approach. They
also acknowledge R.~Ferrini and R.~Houdr\'e for useful discussions
in the preliminary stages of the work concerning possible
experimental realizations of the proposed structure. D.~G.
gratefully acknowledges C.~Ciuti, H.~T\"{u}reci, and
A.~Imamo\u{g}lu for insightful discussions, and A.~Badolato,
K.~Hennessy, and M.~Winger for preliminary experimental results.
\end{acknowledgments}


\begin{thebibliography}{99}
\bibitem{note_dimensionality}
A more general condition is that the photon dimensionality be
lower than that of the exciton: see, e.g., the discussion in
Ref.~\onlinecite{andreani_sif}.
\bibitem{note_dead_layer}
We are assuming that the center-of-mass wavefunction can still be
factorized when a photonic pattern is present. This is true
provided the relative wavefunction is not modified when the
exciton approaches the air holes, i.e., when dead-layer effects
can be ignored. The assumption is reasonable as the exciton Bohr
radius is much smaller than the typical scale of the photonic
lattice.
\bibitem{note_commutator}
Notice that in the following we are assuming $\mathbf{A}$ and
$\mathbf{p}$ to commute. Indeed, the additional terms arising from
non-commutativity depend on the gradient
$\nabla\epsilon(\mathbf{r})$ of the dielectric function, which
vanishes inside the dielectric regions. The surface term at the
hole sidewalls can also be dropped, because the exciton
center-of-mass wavefunction vanishes at the hole boundaries in the
limit $V_{\infty}\rightarrow\infty$.
\bibitem{note_phase}
The phases of the fields have to be such that $C_{-\kk n \nu}=C_{\kk
n \nu}^{\ast}$. This condition is fulfilled by the GME
implementation used here.
\bibitem{note_coupling}
We notice that under this approximation the term proportional to
$\mathbf{A}^2$, which gives rise to the hamiltonian reported in
Eq.~\ref{eq:qwham_bis}, has negligible effect on the energy shift of
mixed exciton-photon modes.\cite{savona94} Such term is included
here for completeness, as well as to highlight the similarities
between the present formalism and the original one by
Hopfield\cite{hopfield58} for bulk exciton-polaritons.
\bibitem{note_overlap}
In order to compare Eq. (\ref{eq:integral}) with the
exciton-photon coupling matrix element calculated, e.g., in Refs.
\onlinecite{savona94,panzarini99}, we notice that the expression
for the overlap integral given here has dimensions $1/ \sqrt{L}$,
and it already contains the effects of vertical confinement on the
electric field profile.
\bibitem{note_parameters}
In the numerical calculations shown in the present work,
convergence of photonic modes is achieved by using 101 plane waves
and 2 guided modes in the basis set for GME. The number of exciton
levels taken in the Hopfield matrix is typically
$M_{\mathrm{max}}\leq 10$, which is largely sufficient to reach
convergence for Rabi splitting with the photonic modes considered.
A typical computation takes only a few minutes on a Personal
Computer.
\end{thebibliography}
\end{document}